\documentclass[english,aps,superscriptaddress,nofootinbib,nobibnotes,notitlepage,preprintnumbers,showpacs]{revtex4-1}
\pdfoutput=1
\usepackage{lmodern}
\DeclareMathAlphabet{\mathbbold}{U}{bbold}{m}{n} 

\usepackage[T1]{fontenc}
\usepackage[latin9]{inputenc}
\usepackage{geometry}
\usepackage{color}
\usepackage{babel}
\usepackage{amsmath}
\usepackage{amssymb}
\usepackage{cancel}
\usepackage{graphicx}
\usepackage{esint}
\usepackage{caption}
\usepackage[unicode=true,pdfusetitle,
 bookmarks=true,bookmarksnumbered=false,bookmarksopen=false,
 breaklinks=false,pdfborder={0 0 1},backref=false,colorlinks=true]
 {hyperref}
\hypersetup{
 citecolor=blue,linkcolor=blue,urlcolor=blue}

\makeatletter

 \@ifundefined{textcolor}{}
 {%
   \definecolor{BLACK}{gray}{0}
   \definecolor{WHITE}{gray}{1}
   \definecolor{RED}{rgb}{1,0,0}
   \definecolor{GREEN}{rgb}{0,1,0}
   \definecolor{BLUE}{rgb}{0,0,1}
   \definecolor{CYAN}{cmyk}{1,0,0,0}
   \definecolor{MAGENTA}{cmyk}{0,1,0,0}
   \definecolor{YELLOW}{cmyk}{0,0,1,0}
 }

\usepackage{babel}

\makeatletter
\def\simgt{\mathrel{\lower2.5pt\vbox{\lineskip=0pt\baselineskip=0pt
           \hbox{$>$}\hbox{$\sim$}}}}
\def\simlt{\mathrel{\lower2.5pt\vbox{\lineskip=0pt\baselineskip=0pt
           \hbox{$<$}\hbox{$\sim$}}}}
\makeatother

\numberwithin{equation}{section}

\newcommand{\be}{\begin{equation}}
\newcommand{\bea}{\begin{eqnarray}}
\newcommand{\ee}{\end{equation}}
\newcommand{\eea}{\end{eqnarray}}
\newcommand\beq{\begin{equation}}
\newcommand\eeq{\end{equation}}

\begin{document}

\title{Non-linear Holographic Entanglement Entropy Inequalities for Single Boundary 2D CFT}

\author{Emory Brown}
\affiliation{Institute for Quantum Information and Matter}
\affiliation{Walter Burke Institute for Theoretical Physics, \\California Institute of Technology 452-48, Pasadena, CA 91125, USA}
\author{Ning Bao}
\affiliation{Institute for Quantum Information and Matter}
\affiliation{Walter Burke Institute for Theoretical Physics, \\California Institute of Technology 452-48, Pasadena, CA 91125, USA}
\author{Sepehr Nezami}
\affiliation{Stanford Institute for Theoretical Physics, Stanford University, Stanford, CA 94305, USA}

\begin{abstract}
Significant work has gone into determining the minimal set of entropy inequalities that determine the holographic entropy cone. Holographic
systems with three or more parties have been shown to obey additional inequalities that generic quantum systems do not. 
We consider a two dimensional conformal field theory that is a single boundary of a holographic system and find four additional 
non-linear inequalities which are derived from strong subadditivity and the formula for the entanglement entropy of a region
on the conformal field theory. We also present an equality obtained by application of a hyperbolic extension of Ptolemy's theorem to a
two dimensional conformal field theory.
\end{abstract}

\maketitle

\section{Introduction}
Finding new information theoretic inequalities in quantum information theory has proven to be a difficult enterprise.
Since the discovery of strong subadditivity and weak monotonicity almost fifty years ago
\citep{SSA}, there have been no new non-conditional confirmed
inequalities that are provably true for all quantum systems, even though there is strong evidence that such inequalities must exist \citep{Ibinson}. There has been, however,  recent progress on inequalities that are true for all
quantum systems that possess a semi-classical gravity dual through holography. In particular, it has been shown that such systems satisfy monogamy
of mutual information (MMI), as shown by \citep{HHM}:
\begin{equation}
S_{AB}+S_{BC}+S_{AC}\geq S_A+S_B+S_C+S_{ABC}.
\end{equation}
for systems of three parties or greater. Indeed, for 3 or 4 party holographic systems this new 
relation is the only additional inequality needed to completely parametrize the space of allowed entanglement entropies\citep{Bao}, or the "holographic cone." For holographic systems with more than 4 parties, further true nontrivial holographic entanglement entropy inequalities have been found, as well, including
an infinite family consisting of generalizations of the MMI inequality\citep{Bao}. We will defer to those works for more detailed review of the specific
methods used therein.

It is interesting to note that the inequalities found in \citep{Bao} are all linear inequalities that relate finite sums of entanglement entropies to finite sums of entanglement entropies. It can easily be shown, however, that for the extremal rays that bound the holographic cone are generally realized by multiboundary wormhole geometries.\citep{Worm} The reason for this is that for geometries with only a single CFT 
boundary, it is not possible to simultaneously set the relative ratios of both the finite and divergent parts of the entanglement entropy to the same relative
 ratios. It is therefore possible that if we restrict our attention to the set of holographic theories for which there is only one boundary CFT we will find new, and indeed non-linear entanglement entropy inequalities. Such theories are of great interest in holography, as in many cases analyses of a single CFT are much more tractable than multiple entangled CFT's.

It is useful at this point to briefly review the relevant entanglement entropy equation for conformal field theory. The Cardy entanglement entropy formula
gives, for a two-dimensional CFT, that
\begin{equation}
S_{A}=\frac{c}{3}\ln{\left(\frac{L}{\pi\epsilon}\sin{\frac{\pi A}{L}}\right)}
\end{equation}
where $c$ is the central charge of the CFT, $L$ is the length of
the total system, $A$ is the length of subsystem $A$, and $\epsilon$
is the UV cutoff\citep{CC}. We will focus on the cse where $A<<L$, where the above simplifies to the more familiar
\begin{equation}
S_{A}=\frac{c}{3}\ln{\frac{A}{\epsilon}}.
\end{equation}

In addition to the fact that the above equation exists, two dimensional conformal field theories are interesting 
as the Virasoro algebra describing their symmetries is infinite dimensional, allowing very general statements
to be made about them solely from symmetry considerations. 

The organization of the paper will be as follows: section II will describe the method we use to find new non-linear inequalites of the type described above. Section III will present a novel non-linear equality derivable from the hyperbolic generalization of Ptolemy's theorem. Finally, we conclude in section IV
with some future directions.

\section{Non-linear Entanglement Entropy Inequalities}

Let us consider the case of two regions $A$ and $B$ on a 2-D CFT separated by a region $S$ with all three regions small 
compared to the total size of the CFT. 
\begin{figure}[h!]
\centering
\includegraphics[width=0.5\textwidth]{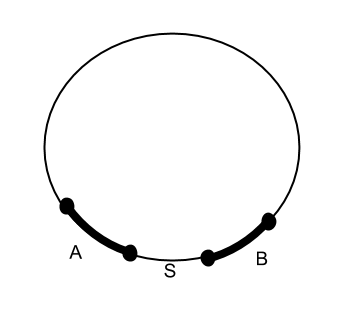}
\caption*{Figure 1: Two regions A and B on a 2D CFT separated by a region S}
\end{figure}
There are two cases for the term $S_{AB}$ depending on
the size of the separation region relative to the primary regions:
\begin{figure}[h!]
\centering
\includegraphics[width=\textwidth]{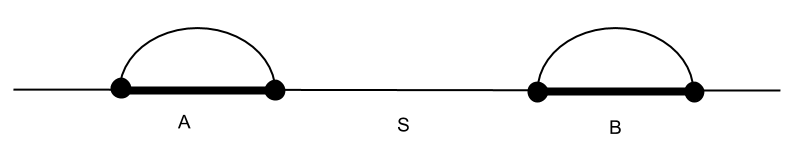}
\caption*{Figure 2a: $S_{AB}=S_{A}+S_{B}$}
~
\includegraphics[width=\textwidth]{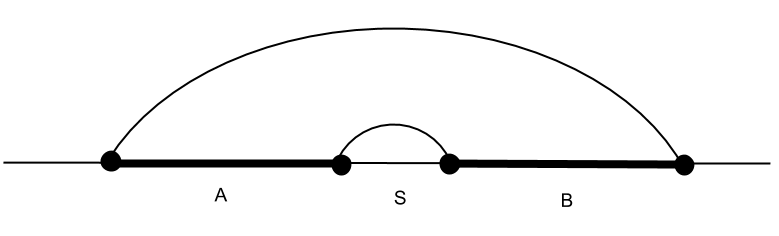}
\caption*{Figure 2b: $S_{AB}=S_{ABS}+S_{S}$}
\end{figure}
In other words, if $S\geq\sqrt{\frac{(A+B)^{2}}{4}+AB}-\frac{A+B}{2}$ then  $S_{AB}=S_{A}+S_{B}$, and if  $S\leq\sqrt{\frac{(A+B)^{2}}{4}+AB}-\frac{A+B}{2}$ then $S_{AB}=S_{ABS}+S_{S}$.

Let's consider the standard subadditivity inequality, 
\begin{equation}
S_A+S_B\geq S_{AB}.
\end{equation}
Let's consider the case where $S_{AB}=S_S+S_{ABS}$. Using Cardy-Calabrese from above we can rewrite this as
\begin{equation}
S_A\geq S_S+\frac{c}{3}\ln{\frac{ABS}{\epsilon}}-\frac{c}{3}\ln{\frac{B}{\epsilon}}.
\end{equation}
By combining the logarithms, we interestingly get that the $\epsilon$'s cancel:
\begin{equation}
S_A\geq S_S+\frac{c}{3}\ln{\frac{ABS}{B}}.
\end{equation}
which becomes
\begin{equation}
S_A\geq S_S+\frac{c}{3}\ln{\left(1+\frac{AS}{B}\right)}.
\end{equation}
Recall that, using Cardy-Calabrese as above, that
\begin{equation}
S_{AS}-S_B=\frac{c}{3}\ln{\frac{AS}{B}},
\end{equation}
which we can invert to get that
\begin{equation}
\frac{AS}{B}=e^{\frac{3}{c}(S_{AS}-S_B)}.
\end{equation}
Inserting this back into equation II.4, we get
\begin{equation}
S_A\geq S_S+\frac{c}{3}\ln{\left(1+e^{\frac{3}{c}(S_{AS}-S_B)}\right)}.
\end{equation}
Using similar methods, we are able to prove four such independent non-linear inequalities, by taking different limits of A, B, and S, and considering as our initial inequality either strong subadditivity or subadditivity. It's interesting to note that beginning with weak monotonicity or Araki-Lieb does not give non-indepedent inequalities from SSA and SA. We report the remaining inequalities in a table below.

\begin{tabular}{|c|c|c|}
\hline
Initial Inequality & Non-linear Inequality & Restrictions\tabularnewline
\hline 
\hline 
$S_{A}+S_{B}\geq S_{AB}$ & $S_{A}\ge S_{S}+\frac{c}{3}\ln{\left(1+e^{\frac{3}{c}(S_{AS}-S_{B})}\right)}$ & $S_{AB}=S_{ABS}+S_{S}$\tabularnewline
\hline 
$S_{AB}+S_{AS}\geq S_{A}+S_{ABS}$ & $S_{AB}\geq S_{A}+\frac{c}{3}\ln{\left(1+e^{\frac{3}{c}(S_{B}-S_{AS})}\right)}$ & \tabularnewline
\hline 
$S_{AS}+S_{BS}\geq S_{S}+S_{ABS}$ & $S_{AS}\geq S_{S}+\frac{c}{3}\ln{\left(1+e^{\frac{3}{c}(S_{B}-S_{AS})}\right)}$ & \tabularnewline
\hline 
$S_{AS}+S_{BS}\geq S_{S}+S_{ABS}$ & $S_{AS}\geq \frac{c}{3}\ln{\left(1+e^{\frac{3}{c}(S_{AS}-S_{B})}\right)}$ & $S_{AB}=S_{A}+S_{B}$\tabularnewline
\hline 
\end{tabular}

These inequalities provide additional nonlinear constraints on top of the known holographic cone constraints that apply when there is only one CFT boundary.

\section{Hyperbolic Extension of Ptolemy's Theorem}

There is another novel constraint that can be made for systems with only a single CFT boundary, resulting from Ptolemy's theorem, and the 
Ryu-Takayanagi \citep{RT} dual description of entanglement entropies as geodesics, that
\begin{equation}
S_A=\frac{Area}{4G_N}
\end{equation}

Recall that Ptolemy's theorem in plane geometry states that, for cyclic quadrilaterals, the product of the lengths of the diagonals is equal to the sum
of the product of the lengths of the opposing sides.

There exists an extension of Ptolemy's theorem to hyperbolic space that states that four points in a space with a Poincar\'{e} metric lie on a circle, line,
horocircle, or one branch of an equidistant curve if and only if the determinant $|\sinh^{2}(P_{i}P_{j}/2)|$ vanishes, where $P_{i}P_{j}$
is the geodesic length between points $P_{i}$ and $P_{j}$. Bao et al. \citep{Bao} argued that for single boundary spacetimes the divergent portion of the geodesic lengths must cancel allowing us to consider only the finite portion of the geodesic lengths. If we consider two disjoint regions $A$ and $B$ on a 2D CFT, then the
boundary points of these regions lie on a circle in the dual $AdS_{3}$ theory. 
\begin{figure}[h!]
\centering
\includegraphics[width=0.5\textwidth]{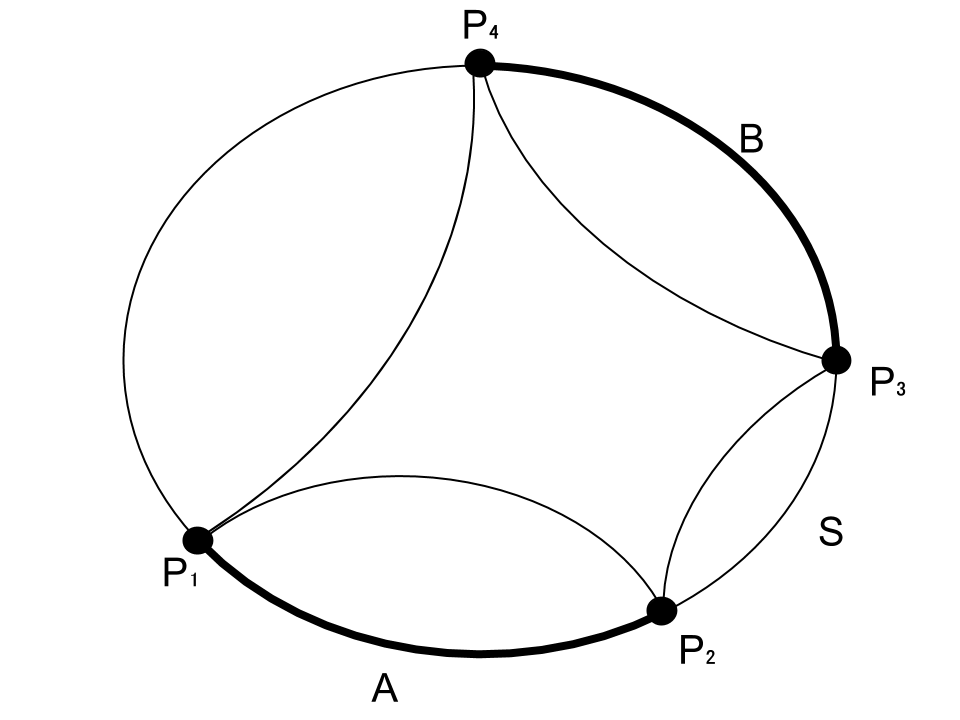}
\caption*{Figure 3: The geodescics in $AdS_{3}$ connecting boundary points of two disjoint regions on a 2D CFT}
\end{figure}
We then know that the determinant $|\sinh^{2}(2G_{N}^{3}S_{ij})|$
vanishes where $S_{ij}$ is the entanglement entropy of the region defined by the boundary points $i$ and $j$. Evaluating this and
expressing all entanglement entropies in terms of the two regions $A$ and $B$ as well as the smaller separation region $S$ and dropping the $G$'s we obtain the following 
constraint on 2D CFT entanglement entropies:
\begin{equation}
\begin{split}
\sinh^{4}(2S_{A})\sinh^{4}(2S_{B})-2\sinh^{2}(2S_{A})\sinh^{2}(2S_{BS})\sinh^{2}(S_{AS})\sinh^{2}(2S_{B}) \\
-2\sinh^{2}(2S_{A})\sinh^{2}(2S_{S})\sinh^{2}(2S_{ABS})\sinh^{2}(2S_{B})+\sinh^{4}(2S_{BS})\sinh^{4}(2S_{AS}) \\
-\sinh^{2}(2S_{BS})\sinh^{2}(2S_{S})\sinh^{2}(2S_{ABS})\sinh^{2}(2S_{AS})+\sinh^{4}(2S_{S})\sinh^{4}(2S_{ABS})=0.
\end{split}
\end{equation}

We note that the equality that we describe above would be turned into an inequality due to corrections from bulk matter density. This can be derived in a more formal way using Gromov $CAT(k)$ space\citep{Sepehr}. Unfortunately this does not appear to be very amenable to simplification, though it does have the pleasing feature of being an equality as opposed to an inequality that is true for two dimensional CFT that are a single boundary of a semi-classical gravitational theory. 

\section{Conclusion}

In this paper we presented four new non-linear entanglement entropy inequalities derived from strong subadditivity and subadditivity as well as one exact
equality derived from a hyperbolic extension of Ptolemy's theorem, all of which hold for regions on a 2D CFT. These findings provide significant additional 
constraints on the entanglement entropies of regions of 2D CFTs of a single boundary.

A potential direction for future investigation is what happens when we extend this strategy to higher dimensional CFT's that are nevertheless also single boundary holographic theories. The arguments that the extremal rays of the full holographic cone cannot be realized on a single boundary are not dependent on dimension. While it is true that the precise form of the entanglement entropy is not known in those theories, because much of the argument is indepedent of an overall scaling it's possible that knowing the scaling of the entanglement entropies would be sufficient to find similar types of inequalities.

\vspace*{\baselineskip}

We thank Sepehr Nezami, Hirosi Ooguri, John Preskill, Bogdan Stoica, James Sully, and Michael Walter.  The work of Ning Bao is supported by a Dubridge Postdoctoral Fellowship.

\end{document}